\documentclass[twocolumn,aps,prl,final,superscriptaddress]{revtex4-1}
\usepackage[latin9]{inputenc}
\usepackage[active]{srcltx}
\usepackage{color}
\usepackage{amsmath}
\usepackage{graphicx}
\usepackage{subscript}
\usepackage[unicode=true,pdfusetitle,
 bookmarks=true,bookmarksnumbered=true,bookmarksopen=true,bookmarksopenlevel=1,
 breaklinks=true,pdfborder={0 0 1},backref=false,colorlinks=true]
 {hyperref}
\hypersetup{
 linkcolor=blue,urlcolor=blue,citecolor=blue,pdfstartview={FitH},hyperfootnotes=false}

\makeatletter

\newcommand*\LyXPunctSpace{\hphantom{,}}

\usepackage{dcolumn}
\usepackage{bm}

\usepackage{times}

\makeatother

\begin{document}
\title{Strain-induced bent domains in ferroelectric nitrides}
\author{Zhijun Jiang}
\email{zjjiang@xjtu.edu.cn}

\affiliation{Ministry of Education Key Laboratory for Nonequilibrium Synthesis
and Modulation of Condensed Matter, Shaanxi Province Key Laboratory
of Advanced Functional Materials and Mesoscopic Physics, School of
Physics, Xi'an Jiaotong University, Xi'an 710049, China}
\affiliation{State Key Laboratory of Surface Physics and Department of Physics,
Fudan University, Shanghai 200433, China}
\author{Zhenlong Zhang}
\affiliation{Ministry of Education Key Laboratory for Nonequilibrium Synthesis
and Modulation of Condensed Matter, Shaanxi Province Key Laboratory
of Advanced Functional Materials and Mesoscopic Physics, School of
Physics, Xi'an Jiaotong University, Xi'an 710049, China}
\author{Charles Paillard}
\affiliation{Smart Ferroic Materials Center, Physics Department and Institute for
Nanoscience and Engineering, University of Arkansas, Fayetteville,
Arkansas 72701, USA}
\affiliation{Université Paris-Saclay, CentraleSupélec, CNRS, Laboratoire SPMS,
91190 Gif-sur-Yvette, France}
\author{Hongjun Xiang}
\email{hxiang@fudan.edu.cn}

\affiliation{Key Laboratory of Computational Physical Sciences (Ministry of Education),
Institute of Computational Physical Sciences, State Key Laboratory
of Surface Physics and Department of Physics, Fudan University, Shanghai
200433, China}
\author{Laurent Bellaiche}
\email{laurent@uark.edu}

\affiliation{Smart Ferroic Materials Center, Physics Department and Institute for
Nanoscience and Engineering, University of Arkansas, Fayetteville,
Arkansas 72701, USA}
\begin{abstract}
Ferroelectric nitrides have emerged as promising semiconductor materials
for modern electronics. However, their domain structures and associated
properties are basically unknown, despite their potential to result
in optimized or new phenomena. Density functional theory calculations
are performed to investigate the effect of epitaxial strain on multidomains
of (Al,Sc)N nitride systems and to compare it with the monodomain
case. The multidomain systems are predicted to have five strain-induced
regions, to be denoted as Regions I to V, respectively. Each of these
regions is associated with rather different values or behaviors of
physical properties such as axial ratio, polarizations, internal parameters,
bond lengths, etc. Of particular interest is the prediction of bent
domains under compressive strain extending beyond $-$5.5\%, which
indicates that domain walls may play a key role in the mechanical
failure properties of these systems. Interestingly, such bending induces
the creation of a finite in-plane polarization (in addition to out-of-plane
dipoles) due to geometric and symmetry considerations. Strikingly
too, the bent domains have lower energy than the wurtzite monodomains
and have atomically sharp boundaries. Our findings may pave the way
for domain wall engineering in ferroelectric nitrides. 
\end{abstract}
\maketitle

\section{Introduction}

Wurtzite AlN is a pyroelectric material but is not ferroelectric since
its polarization direction cannot be switched by an electric field
\citep{Wright1995,Bungaro2000}, which limits its applications \citep{Bernardini1997,Dreyer2016}.
On the other hand, new wurtzite-type ferroelectrics have been achieved
by doping it with Sc to make Al$_{1-x}$Sc$_{x}$N compounds \citep{Fichtner2019},
for which the polarization-\textit{versus}-electric field hysteresis
loops were observed for different Sc compositions. Such promising
nitride system for technological applications \citep{Dawber2005,Scott2007}
has therefore attracted great attention. In particular, its spontaneous
polarization, that lies along the $z$-axis, can be larger than 1.0
C/m$^{2}$, enabling applications for random access memories, actuators,
and sensors \citep{Yasuoka2020,Yazawa2021,Wolff2021,Wang2021,Tsai2021,Guido2023,Wang2023,Zheng2023,Guido2023-1}.

Numerous theoretical studies have also been conducted in (Al,Sc)N
alloys \citep{Tasnadi2010,Zhang2013,Talley2018,Wang2021-1,Zang2023,Hwang2024}
and AlN/ScN superlattices \citep{Alam2019,Jiang2019,Jiang2021,Ye2021},
to explain the origin of the ferroelectricity, as well as to predict
and understand other physical properties, including piezoelectricity,
electro-optic conversion and energy storage. Of particular importance
for these properties is the existence of a nonpolar metastable hexagonal
phase in ScN~\citep{Farrer2002,Ranjan2003} and the possibility to
continuously go from such nonpolar state to the wurtzite structure
via a physical handle, that may be epitaxial strain or composition.

Moreover, ferroelectric materials typically exhibit domains, which
are regions of uniform polarization usually forming during the paraelectric--to--ferroelectric
phase transition~\citep{Damjanovic1998}. The resulting domain wall
separating two domains can adopt novel or enhanced characteristics,
such as photovoltaic effects \citep{Alexe2011}, electronic conductivity
at ferroelectric domain walls \citep{Seidel2009,Farokhipoor2011}
and strong magnetoresistance in BiFeO$_{3}$ domain walls \citep{He2012},
some of which potentially mediated by the attraction of defects by
the domain wall \citep{Paillard2017,Cheng2023}. All these singular,
localized, properties are expected to lead to novel applications in
nanoelectronics \citep{Catalan2012}. To the best of our knowledge,
domains and their possible evolutions with epitaxial strains are not
well known in (Al,Sc)N systems. One may in fact wonder if surprises
are in store there, especially when also realizing that polarization
switching was found to occur via a novel mechanism in nitride ferroelectrics
\citep{Zhu2021,Calderon2023}.

In the present study, we conduct first-principles calculations to
investigate the evolution of strain-induced properties in multidomains
and monodomains made of AlN/ScN superlattices. We do reveal the existence
of unusual features, including multidomains that (1) are bent and
that give rise to an additional in-plane polarization, that superimposes
on typical out-of-plane dipoles; and (2) have atomically sharp domain
walls.

This article is organized as follows. Section II provides details
about the methods. Sections III presents properties of the multidomain
and monodomain. Finally, Sec. IV gives a summary of this work.

\section{Methods}

Two types of $1\times1$ AlN/ScN superlattices structures are initially
considered in the present study. One is a wurtzite monodomain with
a spontaneous polarization along the $z$-direction. The other one
is a wurtzite multidomain with 180$^{\circ}$ uncharged domain walls.
Practically, the periodic supercell of the multidomain structure consists
of two domains separated by two domain walls. As illustrated in Fig.~\ref{fig:structure}(c),
the polarization in the left and right domains in the initial multidomain
structure is along the $+z$ and $-z$ direction of the supercell,
respectively. These two supercell contain 12 $\times$ 1 $\times$
2 unit cells that possess 96 atoms and that are periodic along the
$x$, $y$, and $z$ axes. Note that head-to-head and tail-to-tail
domain structures are, in the absence of free carriers to screen the
polarization bound charges, unstable in $1\times1$ AlN/ScN superlattices.

First-principles calculations are performed on the multidomain and
monodomain structures based on the density functional theory (DFT)
with the generalized gradient approximation of the Perdew-Burke-Ernzerhof
(PBE) exchange-correlation functional form \citep{Perdew1996}, using
the Vienna \textit{ab initio} simulation package (VASP) \citep{Kresse1996-1,Kresse1996}.
The projector augmented-wave (PAW) method \citep{Blochl1994} implemented
in VASP is employed to treat the interaction between core and valence
electrons. A $\Gamma$-centered 1 $\times$ 11 $\times$ 3 $k$-point
mesh is used to sample the Brillouin zone of the multidomain and monodomain
and a plane-wave cutoff of 500 eV is adopted for all calculations.
Different in-plane lattice constants, $a$, are chosen and then kept
fixed while the out-of-plane lattice vector is allowed to relax in
the simulations, and the atomic positions are fully optimized until
the ionic forces are less than 0.001 eV/Å. The epitaxial strain is
defined as $\eta_{\textrm{in}}=(a-a_{\textrm{eq}})/a_{\textrm{eq}}$,
where $a_{\textrm{eq}}$ is the in-plane lattice constant corresponding
to the \textit{equilibrium} structure of the monodomain in AlN/ScN
superlattices, contrary to Ref.~\citep{Jiang2019} that chose the
hexagonal nonpolar structure as the reference for 0\% strain. In the
present study, the considered resulting strains are ranging between
$-$10\% and $+$10\%. Note that an epitaxial compressive strain of
about $-$7\% and a tensile strain exceeding $+$8\% were experimentally
realized in BiFeO$_{3}$ thin films \citep{Zeches2009,Sando2016}
and La$_{0.7}$Ca$_{0.3}$MnO$_{3}$ membranes \citep{Hong2020}.
Note also that our monodomain relaxed structure is in good agreement
with previous theoretical results \citep{Alam2019}. For completeness,
we also considered rocksalt and zincblende structures with chemical
ordering along the $[111]$ cubic direction (see Fig.~S1 of the Supplementary
Material (SM) \citep{SM}).

\section{Results and discussion}

\subsection{Structural properties in multidomain and monodomain }

\begin{figure}
\includegraphics[width=9cm]{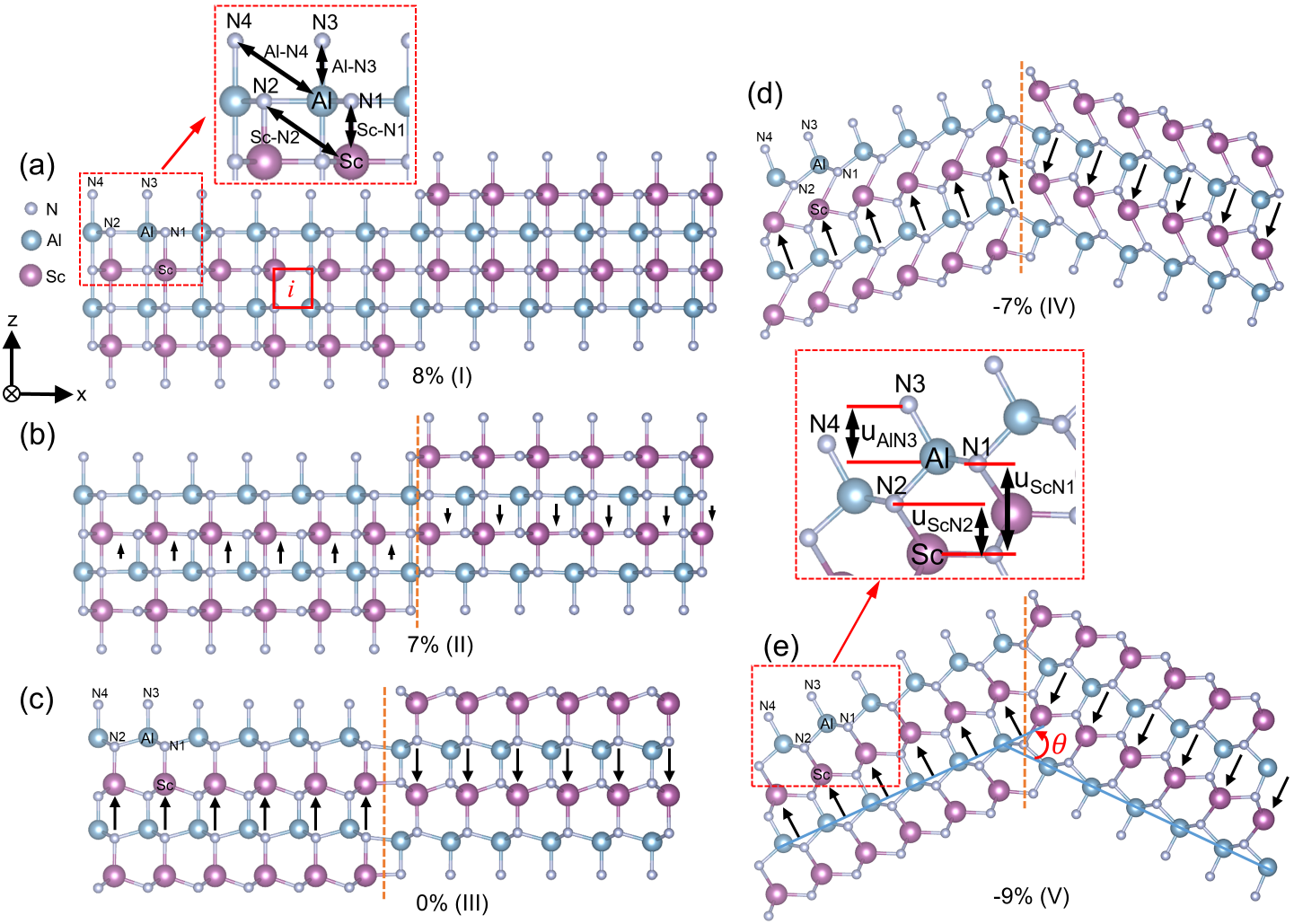}

\caption{Crystal structures of AlN/ScN superlattices. Panel (a) depicts the
nonpolar hexagonal-derived structure in Region I for an epitaxial
strain being equal to $+$8\% (the red solid box defines a unit cell
at number $i$). Panel (b) shows an intermediate multidomain structure
of Region II for strain at $+$7\%. Panel (c) displays a wurtzite-derived
multidomain configuration in Region III for $\eta_{{\rm in}}$ $=$
0\%. Panel (d) shows the multidomain structure at $-$7\% strain in
Region IV. Panel (e) depicts the multidomain structure at $-$9\%
strain in Region V. The arrows represent the direction of the electric
dipoles. The orange dashed lines represent the position of the domain
walls. The bond lengths Sc-N1, Sc-N2, Al-N3 and Al-N4, internal parameters
$u_{\textrm{ScN1}}$, $u_{\textrm{ScN2}}$ and $u_{\textrm{AlN}3}$,
and the definition of the angle $\theta$ are shown in panels (a)
and (e) and their inset. \label{fig:structure}}
\end{figure}

\begin{figure}
\includegraphics[width=8cm]{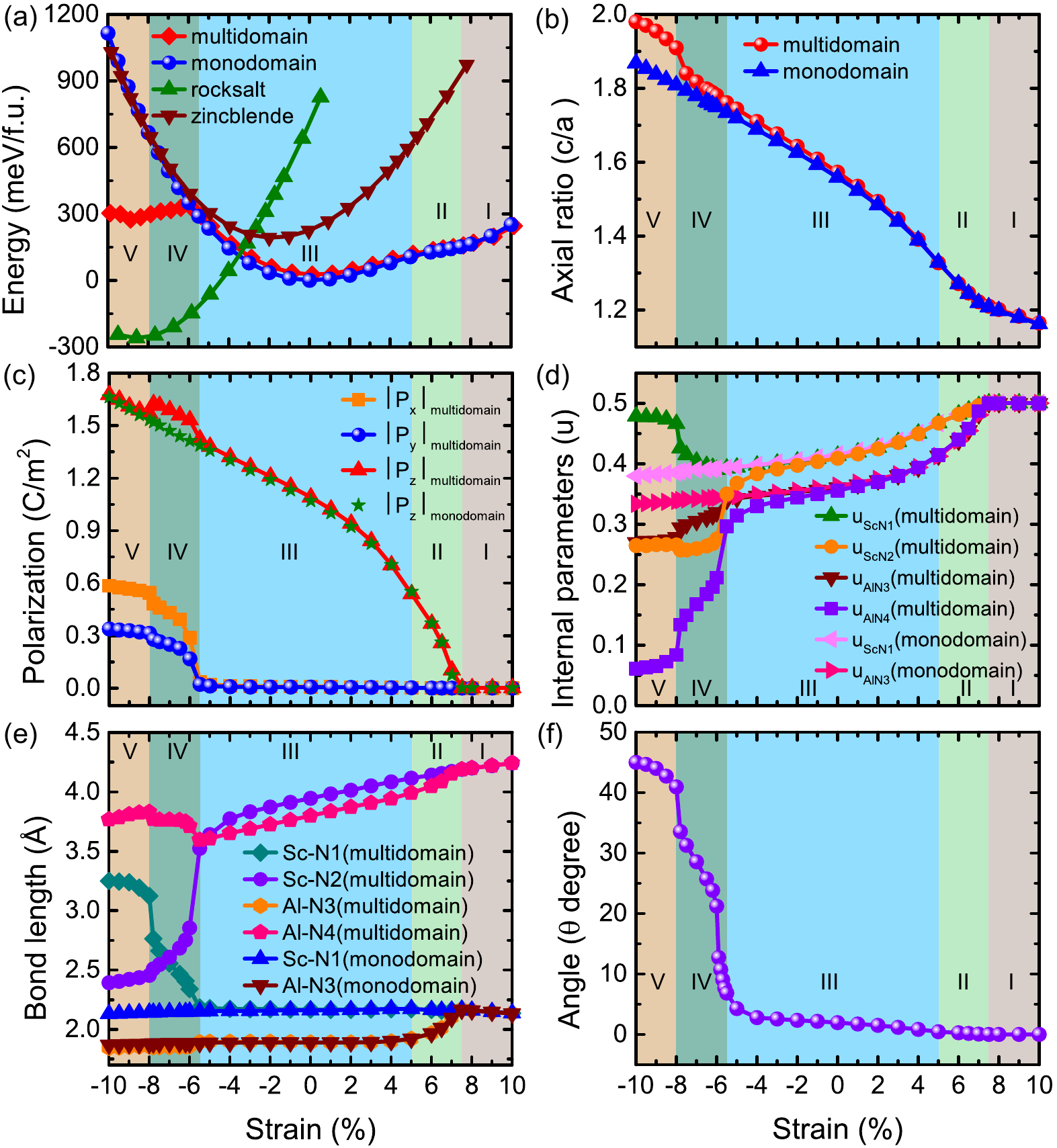}

\caption{Structural properties of AlN/ScN superlattices as a function of strain,
for strains ranging between $-$10\% and $+$10\%: (a) the total energy
of multidomain, monodomain, rocksalt, and zincblende structures in
$1\times1$ AlN/ScN superlattices; (b) axial ratio $c$/$a$ of multidomain
and monodomain; (c) the average of magnitude of the polarization in
multidomain and monodomain; (d) internal parameters $u_{\textrm{ScN}}$
(including $u_{\textrm{ScN1}}$and $u_{\textrm{ScN2}}$) and $u_{\textrm{AlN}}$
(including $u_{\textrm{AlN3}}$and $u_{\textrm{AlN4}}$) of multidomain
and monodomain in AlN/ScN superlattices, respectively; (e) bond lengths
of Sc-N (including Sc-N1 and Sc-N2) and Al-N (including Al-N3 and
Al-N4) in multidomain and monodomain; and (f) angle $\theta$ of multidomain.\label{fig:properties}}
\end{figure}

Figures~\ref{fig:structure}(a)-\ref{fig:structure}(e) show the
relaxed crystal structures of the multidomain for different strains.
Moreover, Figs.~\ref{fig:properties}(a)-\ref{fig:properties}(e)
display different properties of both the multidomain and monodomain
as a function of strain, that are: the total energy (per unit cell)
in our considered AlN/ScN superlattices; the $c$/$a$ axial ratio;
the supercell averaged \textit{magnitude} of the three Cartesian components
of the local polarization ($\left|P_{x}\right|$, $\left|P_{y}\right|$,
and $\left|P_{z}\right|$, where $|P_{\alpha}|=\sum_{i}|P_{i,\alpha}|/N$,
where $\alpha=x$, $y$ or $z$ and where $i$ runs over all the $N$
unit cell of the supercells); different internal parameters $u_{\textrm{ScN}}$
(respectively, $u_{\textrm{AlN}}$) connecting the relative position
along the $z$-axis of Sc (respectively, Al) and some N atoms {[}see
Fig.~\ref{fig:structure}(e){]}; bond lengths between Sc and different
N atoms, as well as between Al and various N atoms {[}see Fig.~\ref{fig:structure}(a){]}.
Furthermore, Fig.~\ref{fig:properties}(f) shows the $\theta$ angle
that is depicted in Fig.~\ref{fig:structure}(e), and that is related
to the deviation of atomic lines from horizontality between the two
domains in the relaxed multidomain configuration. The strain-induced
behaviors of the properties for the multidomain case displayed in
Fig.~\ref{fig:properties} allow the determination of five different
strain regions, that we denote as Regions I to V, respectively.

Region I occurs for tensile strain larger than $+$7.5\% and corresponds
to a layered \textit{nonpolar} hexagonal-derived structure, as evidenced
by $c$/$a$ being close to 1.2, $\left|P_{x}\right|$$=$$\left|P_{y}\right|$$=$$\left|P_{z}\right|$
being zero and the internal parameters being equal to 0.5, as shown
in Figs.~\ref{fig:properties}(b), \ref{fig:properties}(c) and \ref{fig:properties}(d),
respectively. Note that the phase in Region I has the nonpolar $P\bar{6}m2$
space group for both the multidomain and monodomain that have similar
structure {[}see an example of this structure in Fig.~\ref{fig:structure}(a){]},
which also explains why the total energies and $c/a$ axial ratio
of the multidomain and monodomain cases are basically identical there.
The Sc-N1 and Al-N3 bond lengths of Fig.~\ref{fig:properties}(e)
are equal to each other and very slightly increase with strain from
2.14 to 2.17 $\textrm{Å}$ for strains decreasing between $+$10\%
and $+$7.5\%. Similarly, the Sc-N2 and Al-N4 bond lengths, while
also equal to each other, decrease with strain (from 4.24 to 4.19
$\textrm{Å}$) in Region I. Figure~\ref{fig:properties}(f) indicates
that the angle $\theta$, which describes how the two domains are
bent, is null in the entire Region I, as consistent with the nonpolar
hexagonal-derived phase.

The next region, that is Region II, pertains to strain ranging between
$+$7.5\% and $+$5\%. It has the polar $Pmc2_{1}$ space group for
the multidomain \textit{versus} polar $P3m1$ for the monodomain.
It is characterized by the occurrence of a polarization aligned parallel
or antiparallel to the $z$-axis in each domain in the multidomain
(while the polarization is only along $+z$ in the monodomain case),
with $\left|P_{z}\right|$ rapidly increasing from zero to 0.54 C/m$^{2}$
when the strain decreases---as shown in Fig.~\ref{fig:properties}(c).
The $c/a$ axial ratio concomitantly increases from 1.21 to 1.33 in
both the multidomain and monodomain. Meanwhile, the internal parameters
$u_{\textrm{AlN3}}$ and $u_{\textrm{AlN4}}$ of the multidomain both
decrease from 0.5 to 0.41 when reducing the strain from $+$7.5\%
to $+$5\% in Region II, and are now distinct from $u_{\textrm{ScN1}}=u_{\textrm{ScN2}}$
of the multidomain---with $u_{\textrm{ScN1}}=u_{\textrm{ScN2}}$
decreasing from 0.5 to 0.47. Concurrently, the Al-N3 bond length decreases
from 2.17 to 1.92 $\textrm{Å}$ while that of Sc-N1 remains constant
and close to 2.17 $\textrm{Å}$. Moreover, the bond lengths of Sc-N2
(from 4.19 to 4.11 $\textrm{Å}$) and Al-N4 (from 4.19 to 3.99 $\textrm{Å}$)
begin to be distinct from each other and both decrease with strain
when the tensile strain reduces in Region II. Interestingly, the angle
$\theta$ starts to very slightly deviates from zero and weakly increases
from 0 to 0.5 degree. Note that the behaviors of the total energy,
$c/a$, $\left|P_{z}\right|$, internal parameters, and bond lengths
of the monodomain are very close to those of the multidomain in Region
II {[}see Figs.~\ref{fig:properties}(a)-\ref{fig:properties}(e){]}.
Structurally, Region II shows in Fig.~\ref{fig:structure}(b), both
for the monodomain supercell and the domains of the multidomain, a
phase that is in-between the hexagonal-derived nonpolar structure
{[}see Fig.~\ref{fig:structure}(a){]} and the wurtzite-derived structure
{[}see Fig.~\ref{fig:structure}(c){]}.

Region III possesses a strain ranging between $+$5\% to $-$5.5\%.
As in Region II, it holds a $Pmc2_{1}$ space group for the multidomain
and $P3m1$ for the monodomain. It has very small in-plane polarization
components $\left|P_{x}\right|$ and $\left|P_{y}\right|$ (close
to zero), while $\left|P_{z}\right|$ is still nonzero {[}see Fig.~\ref{fig:properties}(c){]}
(as in the ideal wurtzite structure) and increases when decreasing
the strain. The $c$/$a$ axial ratio strengthens from 1.33 to 1.76
for the multidomain (respectively, from 1.33 to 1.73 for the monodomain)
when the strain varies from $+$5\% to $-$5.5\%. The internal parameters
remain equal to each other in the multidomain and monodomain cases,
with the exception of $u_{\textrm{ScN2}}$ and $u_{\textrm{AlN4}}$
close to $-$5.5\% strain (note that $u_{\textrm{ScN2}}$$=$$u_{\textrm{ScN1}}$
and $u_{\textrm{AlN4}}$$=$$u_{\textrm{AlN3}}$ in the monodomain).
All internal parameters decrease when going toward more compressive
strain within Region III. The $\theta$ angle slightly increases from
0.5 to 2.7 degree, making the two corresponding atomic lines bending
more and more from each other, when decreasing the strain from $+$5\%
to $-$4\%. It then rapidly increases from 2.7 to 6.8 degrees for
strains varying between $-$4\% and $-$5.5\%. The behavior of the
internal parameters coupled with that of the $\theta$ angle in Region
III make the Sc-N1 and Al-N3 bond lengths almost constant (around
2.17 and 1.89 $\textrm{Å}$ for Sc-N1 and Al-N3, respectively) in
the multidomain case. Note that Regions II and III are distinct from
each other due to the bond length of Al-N3 significantly decreasing
with strain in Region II while being basically a constant in Region
III. The Sc-N2 and Al-N4 bond lengths decrease significantly from
4.12 to 3.53 $\textrm{Å}$ and from 3.99 to 3.59 $\textrm{Å}$, respectively,
in Region III for the multidomain case, because N2 and N4 atoms are
moving toward Sc and Al atoms, respectively, when the compressive
strain strengthens in magnitude. Structurally speaking, each domain
in the multidomain case (as well as the monodomain) corresponds to
a wurtzite-derived phase in Region III. Note that, in Region III,
the total energy of the multidomain remains higher than that of the
monodomain, as expected. As indicated above, the $c/a$ axial ratio,
$\left|P_{z}\right|$, internal parameters and bond lengths of the
monodomain are very close to the multidomain case in Region III, with
the exception of Sc-N2 and Al-N4 at the boundary of $-$5.5\% (not
shown here). Note also that the system found a minimum in energy in
Region III for both the multidomain and monodomain for 0\% of epitaxial
strain, while characteristics of the ideal wurtzite structure (namely,
$c/a$ close to 1.633 and the internal parameter averaged between
$u_{\textrm{ScN1}}$ and $u_{\textrm{AlN3}}$ being around 0.375)
occur for a strain around $-$2.3\% for both the multidomain and monodomain
supercells.

Moreover, Region IV extends from $-$5.5\% to $-$8\%. Strikingly,
the total energy of the multidomain in this region has now a lower
energy than the wurtzite-like monodomain, implying that the multidomain
adopts a configuration that is more stable here. The $c/a$ axial
ratio is much larger than that of the ideal wurtzite structure (which
is 1.633) and varies from 1.76 to 1.91 for strain ranging between
$-$5.5\% and $-$8\% in the multidomain ($c/a$ increases from 1.73
to 1.82 for the monodomain case). Remarkably, the electric dipoles
are tilted away from the {[}0001{]} direction ($z$-axis) of the ideal
wurtzite structure in each domain of the multidomain case, while the
dipoles of the monodomain are still all along the $z$-axis (as in
the wurtzite structure). Consequently, the multidomain now possesses
both finite $\left|P_{x}\right|$, $\left|P_{y}\right|$, in addition
to significant $\left|P_{z}\right|$. All these three components increase
when strengthening the magnitude of the compressive strain. In the
multidomain, the internal parameter $u_{\textrm{ScN1}}$ increases
with the magnitude of strain while $u_{\textrm{ScN2}}$, $u_{\textrm{AlN3}}$
and $u_{\textrm{AlN4}}$ decrease. These features are due to the fact
that the N1 atom is moving away from Sc while N2, N3 and N4 atoms
are gradually approaching Sc and Al atoms, respectively {[}see an
example of this structure in Fig.~\ref{fig:structure}(d){]} (Note
that, in contrast, the internal parameters of the monodomain all linearly
decrease with strain for that range {[}see Fig.~\ref{fig:properties}(d){]}).
The internal parameters of the multidomain are thus markedly different
from those of the monodomain in Region IV. The aforementioned motions
of the N2 and N4 atoms results in the Sc-N2 bond length decreasing
significantly from 3.53 to 2.45 $\textrm{Å}$ while Al-N4 slightly
increases from 3.59 to 3.83 $\textrm{Å}$ in Region IV for the multidomain.
Furthermore, for the multidomain, the Sc-N1 bond length dramatically
increases from 2.18 to 3.12 $\textrm{Å}$ when increasing the magnitude
of the compressive strain in Region IV {[}see Fig.~\ref{fig:properties}(e){]},
which is due to the combined behavior of $c/a$ and the internal parameter
$u_{\textrm{ScN1}}$. The Sc-N2 bond length then becomes smaller than
that of Sc-N1 for some strains in Region IV. Furthermore, the $\theta$
angle significantly increases from 6.8 to 33.5 degrees when increasing
the amount of compressive strain in Region IV, revealing that atomic
lines between the two domains are strongly bent with respect to each
other {[}see Fig.~\ref{fig:structure}(d){]}. In effect, one Sc atomic
layer appears to slide to allow a reconfiguration of the atomic bonding
within the domain walls with large compressive strain {[}compare Figs.~\ref{fig:structure}(d)
and \ref{fig:structure}(e){]}. Region IV thus acts as a transition
region through which Sc atoms are breaking their bond with N1 to create
a new bond with N2. The stress-strain curve (see Fig.~S2 of the SM
\citep{SM}) shows that the multidomain structure in-plane stress
exhibits a sharp fall, typical of mechanical failure calculated for
instance in nitride monolayers \citep{Mortazavi2022,Ding2016} and
here occurring at $-$5.5\% critical strain. There is thus a strong
possibility that domain walls lower the ability of ferroelectric nitrides
to sustain large elastic deformations. It is interesting to realize
that the bending between two domains was also experimentally and recently
found in 2D van der Waals ferroelectrics such as the $\alpha$-In$_{2}$Se$_{3}$
material \citep{Han2023}, with a bending angle $\theta$ above $\sim$33
degree. Note that $\left|P_{x}\right|$ likely naturally results from
the geometrical bending occurring the $(x,z)$-plane. Mirror symmetry
imposed by the $Pmc2_{1}$ space group further imposes the appearance
of $\left|P_{y}\right|$, with the constraint that $P_{y}$ equals
to $P_{x}/\sqrt{3}$, as demonstrated in Fig.~S3 of the SM \citep{SM}.
Moreover, the structure in each domain of the multidomain corresponds
to a distorted-pyramidal structure because the Sc-N1 and Sc-N2 bond
lengths are very different from those in the case of the wurtzite
configuration. Note that the overall phase in Region IV retains the
$Pmc2_{1}$ space group for the multidomain while the monodomain also
keeps its $P3m1$ space group, as in Regions II and III.

Finally, Region V occurs for strains ranging between $-$8\% and $-$10\%,
with the multidomain and monodomain also preserving their $Pmc2_{1}$
and $P3m1$ space group, respectively. The multidomain has now a much
lower energy than the monodomain {[}see Fig.~\ref{fig:properties}(a){]}.
The boundary between Regions IV and V occurs at a strain of $\sim$
$-$8\%, where the axial ratio $c/a$, the three components of the
polarization, internal parameters, bond lengths, and $\theta$ angle
of the multidomain all experience a significant jump. These jumps,
as well as the retaining of the $Pmc2_{1}$ space group, characterize
a strain-induced first-order \textit{isostructural} transition in
the multidomain supercell (see Refs. \citep{Bellaiche1996,Ranjan2005}
and references therein). The resulting structure in each domain is
a pyramidal structure {[}see Fig.~\ref{fig:structure}(e){]}. The
overall structure in Region V {[}see Fig.~\ref{fig:structure}(e){]}
differs from Region IV {[}see Fig.~\ref{fig:structure}(d){]} because
the $c/a$, internal parameters, bond lengths, and $\theta$ angle
are very different in these two regions in the more stable multidomain
supercell. As a matter of fact, Fig.~\ref{fig:properties}(b) shows
that the $c/a$ axial ratio of the multidomain is extremely large,
and larger than that of the monodomain in Region V. It nearly linearly
increases from 1.91 to 1.98 when strengthening the magnitude of the
compressive strain. The $\theta$ angle concomitantly nonlinearly
increases with strain, adopting enormous values ranging from 40.9
to 45.0 degrees. Consequently, $\left|P_{x}\right|$ (from 0.55 to
0.59 C/m$^{2}$) and $\left|P_{y}\right|$ (from 0.31 to 0.34 C/m$^{2}$)
all nearly linearly increase with strain for the multidomain. $\left|P_{z}\right|$
also increases from 1.57 to 1.67 C/m$^{2}$ in the multidomain (to
be compared from 1.53 to 1.66 C/m$^{2}$ for the monodomain) as a
response to in-plane compressive strain becoming bigger in magnitude.
Furthermore, the internal parameter $u_{\textrm{ScN1}}$ of the multidomain
increases with strain as well (from 0.466 to 0.478) while $u_{\textrm{AlN3}}$
concomitantly slightly decreases (from 0.278 to 0.270) in Region V.
$u_{\textrm{ScN2}}$ and $u_{\textrm{AlN4}}$ of the multidomain also
slightly decrease from 0.266 to 0.265 and from 0.066 to 0.061, respectively,
in Region V. Note that $u_{\textrm{AlN4}}$ is really small, indicating
that the $z$-component of the N4 atomic position is very close to
that of the Al atom in Region V. Figure~\ref{fig:properties}(e)
shows that the bond length of Sc-N1 and Al-N3 slightly increases (from
3.12 to 3.25 $\textrm{Å}$) and decreases (from 1.86 to 1.84 $\textrm{Å}$)
with strain, respectively, in Region V for the multidomain (the bond
length of Sc-N2 and Al-N4 linearly decreases from 2.45 to 2.39 and
from 3.83 to 3.77 $\textrm{Å}$, respectively). Note that the $c/a$,
polarization, internal parameters, and bond lengths behaviors of the
monodomain are very different from those of the multidomain case in
Region V because the monodomain is wurtzite-like while it is a pyramidal
structure for each domain of the multidomain. Note also that the total
energy of the multidomain has a slight minimum around $-$9\%. The
multidomain therefore adopts a metastable state around $-$9\%, in
addition to its wurtzite-like ground state for 0\% strain. Such features
bear resemblance with BiFeO$_{3}$ films, that possess two polymorphs,
usually denoted as the T and R phases, with minima located at very
different strains \citep{Sando2016-1}. It is also worth noting that
the SM \citep{SM} indicates that the stress-strain curves in Region
V recover a linear dependency, indicative of an elastic deformation
regime.

\subsection{Local polarizations in multidomain and monodomain }

\begin{figure}
\includegraphics[width=8cm]{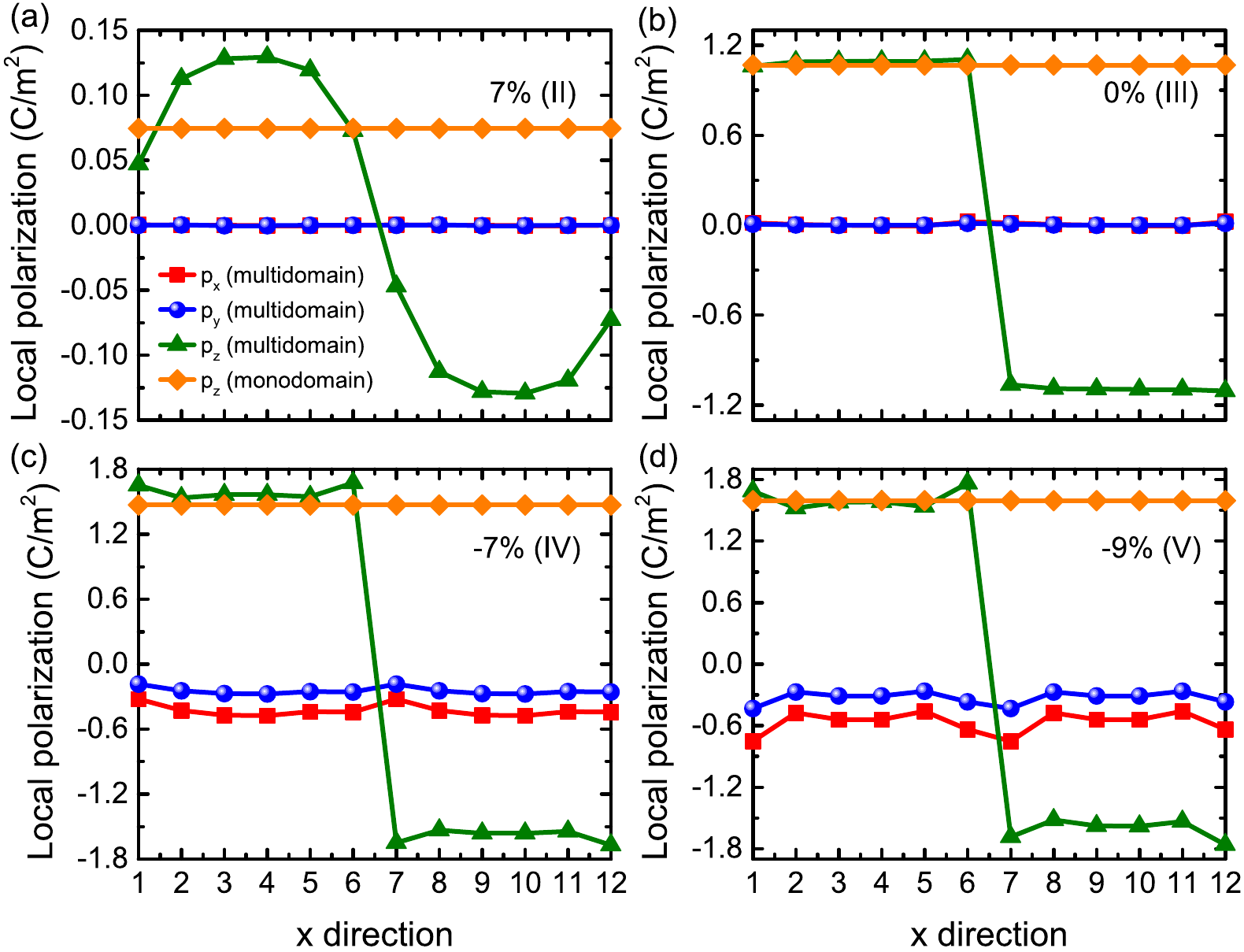}

\caption{The local $x$-, $y$-, $z$-components of polarizations in each unit
cell along the $x$-axis direction for different strains at (a) 7\%
(Region II); (b) 0\% (Region III); (c) $-$7\% (Region IV); and (d)
$-$9\% (Region V) in the multidomain and monodomain cases of AlN/ScN
superlattices, respectively. \label{fig:polarization}}
\end{figure}

Let us now present the evolution of the polarization \textit{plane-by-plane}
for Regions II, III, IV and V in the multidomain case, and also compare
it with the monodomain case. The local polarization in the unit cell
$i$ {[}as shown in Fig.~\ref{fig:structure}(a) by the solid box{]}
is computed via ${\bf p^{\mathit{i}}}=\frac{e}{\Omega}\sum_{j}Z_{j}^{*}{\bf u_{\mathit{j}}}$
\citep{Meyer2002,Neaton2005}, where $e$ is the electron charge,
$\Omega$ is the unit cell volume, $Z_{j}^{*}$ is the Born effective
charge tensor, and ${\bf u_{\mathit{j}}}$ is the atomic displacement
from the ideal lattice site of ion $j$---with $j$ running over
all atoms in the unit cell $i$. Note that Fig.~\ref{fig:properties}(c)
displays the averaged \textit{magnitude} of components of the local
polarization, $\left|P_{x}\right|$, $\left|P_{y}\right|$, $\left|P_{z}\right|$,
while Fig.~\ref{fig:polarization} characterizes \textit{components}
of the plane-by-plane polarizations $p_{x}$, $p_{y}$, $p_{z}$,
that are along the $x$-, $y$- and $z$-axes, respectively. In particular,
only $p_{z}$ is finite in the monodomain and it is homogeneous for
all Regions II-V. Figure~\ref{fig:polarization} indicates that $p_{z}$
\textit{averaged over the entire supercell} is always zero in the
multidomain for all regions because of the opposite directions of
the local out-of-plane polarization in the two domains. Note also
that we do not show the local polarization distributions in Region
I since the phase there is nonpolar because of its hexagonal-derived
structure {[}see Fig.~\ref{fig:structure}(a){]}.

As representative of Region II, we chose a strain of $+$7\%. Figure~\ref{fig:polarization}(a)
shows that the polarization $p_{z}$ in the multidomain behaves like
a sine function along the $x$ direction, changing signs between the
two domains, and generating smooth waves of polarization in Region
II. This is characteristic of Ising-like feature \citep{Gu2014} for
the behavior of $p_{z}$. In comparison, in the monodomain, the local
polarization $p_{z}$ is uniform along the $x$ direction and equal
to 0.075 C/m$^{2}$. This value is smaller in magnitude than that
of the $p_{z}$ of the multidomain for cell numbers ranging from 2
to 5 and from 8 to 11, respectively. On the other hand, it is larger
in magnitude than the $p_{z}$ of the multidomain for cell numbers
1 and 7. Interestingly, the magnitude of $p_{z}$ in the multidomain
at cell numbers 6 and 12 (located in the vicinity of the domain wall)
is very close to that of the monodomain. These behaviors make the
averaged magnitude $\left|P_{z}\right|$ of the multidomain and monodomain
being equal to each other, as shown in Fig.~\ref{fig:properties}(c)
for Region II. Note that the local in-plane polarizations, $p_{x}$
and $p_{y}$, are null at any cell in the multidomain in Region II.

We now chose 0\% of strain as a representative of Region III. Figure~\ref{fig:polarization}(b)
shows that $p_{z}$ in each domain of the multidomain is almost constant
in magnitude along the $x$ direction and equal to $\sim$1.1 C/m$^{2}$.
This $p_{z}$ thus appears to behave like a square wave, because of
its change of sign between the two domains, and is equal in magnitude
to the $p_{z}$ of the monodomain case {[}see Fig.~\ref{fig:polarization}(b){]}.
It is striking to thus realize that there is a sudden jump between
the $p_{z}$ of the two different domains, implying that the domain
wall can be extremely small in this system. This dramatically contrasts
with the case of 90$^{\circ}$ domain walls in ferroelectric perovskites
and domain walls in magnetic systems for which they can extend over
a length of typically 15 and 500 nm, respectively \citep{Dennis1974,Venkat2024}.
These characteristics also make the $\left|P_{z}\right|$ of the multidomain
and monodomain essentially equal to each other in Region III {[}see
Fig.~\ref{fig:properties}(c){]}. Moreover and as shown in Fig.~\ref{fig:polarization}(b),
$p_{x}$ and $p_{y}$ continue to vanish in Region III for the multidomain
and monodomain.

In Region IV, for which we use a representative strain of $-$7\%,
Fig.~\ref{fig:polarization}(c) displays a large value in magnitude
for $p_{z}$ ($\sim$1.59 C/m$^{2}$) in the entire multidomain. This
$p_{z}$ also behaves as a square wave, with a sudden jump between
the two domains and forms a sharp domain wall (note that the bent
domains experimentally found in $\alpha$-In$_{2}$Se$_{3}$ material
are also sharp and can be extremely narrow, down to 2 atoms or 4 Å
wide at their narrowest point \citep{Han2023}). Furthermore, the
$p_{z}$ on the monodomain continues to be homogeneous and is now
equal to 1.47 C/m$^{2}$, which is smaller in magnitude than in the
multidomain ($\sim$1.59 C/m$^{2}$). Consequently, the $\left|P_{z}\right|$
in the multidomain is larger than in the monodomain in Region IV,
as shown in Fig.~\ref{fig:properties}(c). Interesting, Fig.~\ref{fig:polarization}(c)
reveals that the in-plane $p_{x}$ and $p_{y}$ are no longer null
in Region IV. They are quasi-homogeneous ($p_{x}$ $\sim$ $-$0.43
C/m$^{2}$ and $p_{y}$ $\sim$ $-$0.25 C/m$^{2}$, respectively).
The homogeneity of $p_{x}$, including when going from one domain
to another one, guarantees that the domains are uncharged.

Finally, we consider a large compressive strain of $-$9\% for Region
V. Figure~\ref{fig:polarization}(c) shows that the $p_{z}$ of the
multidomain is nearly constant in each domain and displays a large
value in magnitude around 1.61 C/m$^{2}$ for cell numbers ranging
from 1 to 12, which is very close to that of the monodomain (1.6 C/m$^{2}$).
Therefore, the $\left|P_{z}\right|$ in the multidomain and monodomain
are nearly equal to each other, as displayed in Region V of Fig.~\ref{fig:properties}(c).
Once again, the domain walls are sharp for $p_{z}$ in the multidomain
as they present a sudden jump from a large positive value to its opposite
when going from the left domain to the right one. Moreover, $p_{x}$
and $p_{y}$ are nonzero and quasi-homogeneous as well in Region V,
albeit with a larger in magnitude than in Region IV ($p_{x}$ is now
around $-$0.6 C/m$^{2}$ and $p_{y}$ around $-$0.3 C/m$^{2}$ at
$-$9\% strain, respectively) for the multidomain. Note that the near
continuity of $p_{x}$ across the two domains implies that the domain
walls in Region V remain uncharged.

\subsection{Domain wall energy and stability of the multidomain state}

\begin{figure}
\includegraphics[width=8cm]{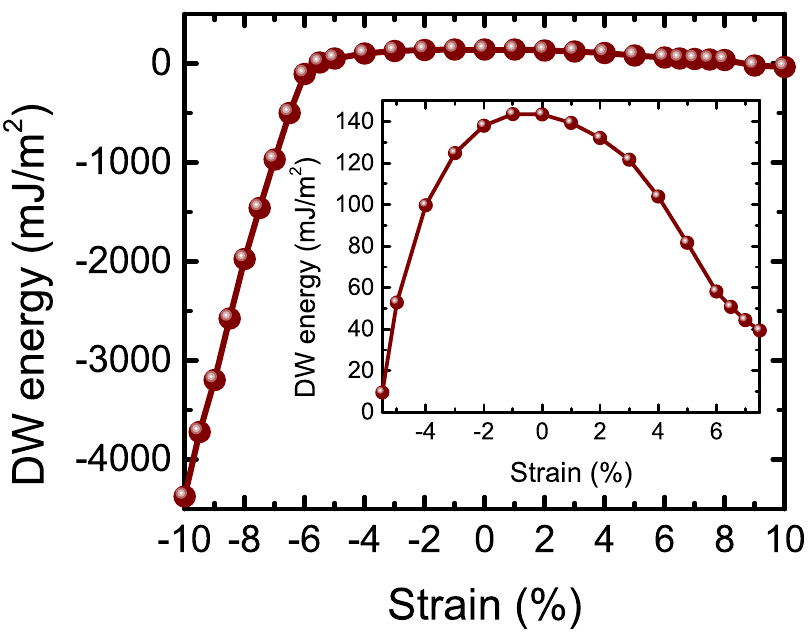}

\caption{Domain wall energy in $1\times1$ AlN/ScN superlattices. The inset
zooms in the data in Regions II and III. \label{fig:dw_energy}}
\end{figure}

The domain wall (DW) energy $E_{\textrm{DW}}$ is computed by the
equation \citep{Hwang2024,Meyer2002,Shimada2008}: $E_{\textrm{DW}}=(E_{\textrm{multi}}-E_{\textrm{mono}})/2S_{\textrm{DW}}$,
where $E_{\textrm{multi}}$ is the total energy of multidomain, $E_{\textrm{mono}}$
is the total energy of monodomain, and $S_{\textrm{DW}}$ is the area
of the domain wall. Figure~\ref{fig:dw_energy} shows the domain
wall energy as a function of strain. For strains ranging between $+$7.5\%
and $+$10\% (Region I), the domain wall energies are close to zero
because both the multidomain and monodomain have a layered \textit{nonpolar}
hexagonal-derived structure there.

The monodomain polar state of $1\times1$ AlN/ScN superlattices is
the most stable state in Regions II and III according to Fig.~\ref{fig:properties}(a).
In these regions, the domain wall energy increases from almost zero
(at the compressive edge of Region III) to 143~mJ/m\textsuperscript{2}
at zero strain (see the inset of Fig.~\ref{fig:dw_energy}). It then
decreases again when one approaches the boundary between Region II
and Region I (nonpolar hexagonal phase). 180 degree domain walls in
our AlN/ScN have thus a comparable surface energy compared to uniaxial
ferroelectrics such as PbTiO\textsubscript{3}, where energies of
the order of 125 mJ/m\textsuperscript{2} were reported~\citep{Paillard2017,Meyer2002}.
Interestingly, both compressive and tensile strain tend to decrease
the domain wall surface energy, likely because phase transitions to
centrosymmetric phases can be expected on both the compressive {[}transition
to the rocksalt phase, see Fig.~\ref{fig:properties}(a){]} and tensile
(transition to the nonpolar hexagonal phase~\citep{Jiang2019}) side.

Interestingly, in the strongly compressive strain regime (Regions
IV and V), the domain wall energies become negative with values ranging
from $-107$ to $-4372$~mJ/m\textsuperscript{2}. The negative domain
wall energies suggest that the bent multidomain state is more energetically
favorable than the monodomain polar state, as consistent with Fig.~\ref{fig:properties}(a).
Interestingly, in Regions IV, where the multidomain state becomes
energetically more favorable with respect to the monodomain state,
marks the onset of mechanical failure, as strain-stress curves (see
Fig.~S2 of the SM \citep{SM}) show a strong drop in the longitudinal
stress facilitated by the strong bending of the domains. Note however,
that elastic behavior is recovered in Region V, indicating that such
bent multidomain state may be metastable at extremely large compressive
strains.

One may wonder whether such bent multidomain state can be accessed
experimentally. Figure~\ref{fig:properties}(a) indicates that this
peculiar state can in fact only be metastable in $1\times1$ AlN/ScN
superlattices. Indeed, in Regions IV and V, the nonpolar rocksalt
phase is numerically found to have lower energy by 300 to 600~meV/f.u.
It is fairly evident that slowly applied strain, such as epitaxial
strain resulting from growth of a thin film on a substrate, could
not address this peculiar domain state. However, large, ultrafast
strain pulses (such as the non-destructive laser shock technique developed
in Ref.~\citep{Deschamp2022}) may be able to address such a metastable
state by creating the necessary amount of deformation to cause mechanical
failure and stabilize the bent domain state. Note also that over AlN/ScN
superlattices, such as $3\times1$, possess a small stability window
where the bent multidomain state is in fact found to be the ground
state (see Fig.~S4 of the SM~\citep{SM} between $-6$ and $-4$~\%
compressive strain). It is thus possible that altering the composition
of the superlattice may reveal such original bent multidomain states.

\section{Summary}

In summary, based on \textit{ab}-\textit{initio} density functional
theory calculations, we predict the existence of five different strain
regions, accompanied by striking structural features, in (uncharged)
multidomains of AlN/ScN superlattices. In particular, for the largest
studied tensile strains, Region I has no local polarization, while
a polarization along $+$$z$ and $-$$z$ develops in each domain
of such multidomain, respectively, in Region II---that exists for
smaller tensile strains and that possesses a sinusoidal behavior of
its plane-by-plane polarization. The polarization in each of these
domains, as well as the axial ratio, significantly grow when going
from tensile to compressive strain in Region III, with the plane-by-plane
polarization now exhibiting a square-like behavior.

Strikingly, these domains bend in Regions IV and V, with this bending
becoming stronger as the compressive strain is enhanced in magnitude
and with the transition from Regions IV to V being isostructural and
of first-order. Such binding results in the appearance of in-plane
and homogeneous polarization. Remarkably, the 180$^{\circ}$ domain
walls in Regions II to V are very sharp. Although these bent domains
do not appear to be stable at thermodynamic equilibrium, it is possible
that strong out-of-equilibrium excitations, such as large strain pulses,
could access this peculiar state which has lower energy than the monodomain
state under large compressive strain. Note that we also checked the
properties of multidomains made of the epitaxial $3\times1$ AlN/ScN
superlattices. They exhibit qualitatively similar results as the presently
studied $1\times1$ AlN/ScN superlattices, in general, and possess
bent domains too, in particular. Interestingly, these bent domains
occur under smaller compressive strain for the $3\times1$ AlN/ScN
superlattices, namely around $-$4\% (see Fig.~S4 of the SM \citep{SM})---with
these bent domains having even lower energy than the rocksalt and
zinc-blende structures, as well as the wurtzite monodomain, for some
strain range. In other words, bent domains can be the ground state
in some cases, and the critical strain at which these intriguing bent
domains appear in ferroelectric nitrides depend on parameters, such
as the overall composition or thickness layers---it is thus also
possible that other superlattice compositions may be able to sustain
such bent domains before the onset of the phase transition to the
centrosymmetric rocksalt structure. 
\begin{acknowledgments}
This work is supported by the National Natural Science Foundation
of China (Grants No.\ 12374092 and No.\ 11804138), Natural Science
Basic Research Program of Shaanxi (Program No.\ 2023-JC-YB-017),
Shaanxi Fundamental Science Research Project for Mathematics and Physics
(Grant No.\ 22JSQ013), ``Young Talent Support Plan'' of Xi'an Jiaotong
University (Grant No.\ WL6J004), the Open Project of State Key Laboratory
of Surface Physics (Grant No.\ KF2023\_06), and the Fundamental Research
Funds for the Central Universities. H.X. acknowledges the support
from the Ministry of Science and Technology of the People's Republic
of China (No. 2022YFA1402901) and NSFC (Grants No.\ 11825403, 11991061,
and 12188101). C.P. and L.B. thank the Defense Advanced Research Projects
Agency Defense Sciences Office (DARPA-DSO) Program: Accelerating discovery
of Tunable Optical Materials (ATOM) under Agreement No. HR00112390142
and the Award No. FA9550-23-1-0500 from the U.S. Department of Defense
under the DEPSCoR program. L.B. also acknowledges the MonArk NSF Quantum
Foundry supported by the National Science Foundation Q-AMASE-i Program
under NSF Award No. DMR-1906383, the ARO Grant No. W911NF-21--1--0113,
and the Vannevar Bush Faculty Fellowship (VBFF) Grant No. N00014-20-1--2834
from the Department of Defense. The HPC Platform of Xi'an Jiaotong
University is also acknowledged. C. P. acknowledges partial support
through Agence Nationale de la Recherche through grant agreement no.
ANR-21-CE24-0032 (SUPERSPIN). 
\end{acknowledgments}

\end{document}